\documentclass[natbib]{emulateapj}
\usepackage{natbib}
\def\s2n{S^{\prime}/N}

\usepackage{amssymb,amsmath}
\def\bs{\boldsymbol}

\slugcomment{Submitted to ApJ, \today}
\shorttitle{Supernova Driving. II. Compressive Ratio in Molecular-Cloud Turbulence}
\shortauthors{Pan et al.}

\begin{document}
\title{Supernova Driving. II. Compressive Ratio in Molecular-Cloud Turbulence}

\author{Liubin Pan}
\affil{Harvard-Smithsonian Center for Astrophysics, 60 Garden Street, Cambridge, MA 02138, USA; lpan@cfa.harvard.edu}
\author{Paolo Padoan,}
\affil{ICREA \& Institut de Ci\`{e}ncies del Cosmos, Universitat de Barcelona, IEEC-UB, Mart\'{i} Franqu\`{e}s 1, E08028 Barcelona, Spain; ppadoan@icc.ub.edu}
\author{Troels Haugb{\o}lle,}
\affil{Centre for Star and Planet Formation, Niels Bohr Institute and Natural History Museum of Denmark, University of Copenhagen, {\O}ster Voldgade 5-7, DK-1350 Copenhagen K, Denmark; haugboel@nbi.ku.dk}
\author{{\AA}ke Nordlund,}
\affil{Centre for Star and Planet Formation, Niels Bohr Institute and Natural History Museum of Denmark, University of Copenhagen, {\O}ster Voldgade 5-7, DK-1350 Copenhagen K, Denmark; aake@nbi.ku.dk}

\begin{abstract}

The compressibility of molecular cloud (MC) turbulence plays a crucial role in star formation models, because it controls the 
amplitude and distribution of density fluctuations. The relation between the compressive ratio (the ratio 
of powers in compressive and solenoidal motions) and the statistics of turbulence has been previously studied systematically only in idealized 
simulations with random external forces. In this work, we analyze a simulation of large-scale turbulence
(250 pc) driven by supernova (SN) explosions that has been shown to yield realistic MC properties. We demonstrate that SN driving
results in MC turbulence with a broad lognormal distribution of the compressive ratio, with a mean value $\approx 0.3$,
lower than the equilibrium value of $\approx 0.5$ found in the inertial range of isothermal simulations with random solenoidal driving. 
We also find that the compressibility of the turbulence is not noticeably affected by gravity, nor are the mean cloud radial (expansion or 
contraction) and solid-body rotation velocities. 
Furthermore, the clouds follow a general relation between the rms density and the rms Mach number similar to that of supersonic 
isothermal turbulence, though with a large scatter, and their average gas density PDF is described well by a lognormal distribution, with 
the addition of a high-density power-law tail when self-gravity is included.

\end{abstract}

\keywords{
ISM: kinematics and dynamics -- MHD -- stars: formation -- turbulence
}

\section{Introduction}

The fragmentation of molecular clouds (MCs) induced by supersonic turbulence is a fundamental aspect of the star formation process.
Recent models of the star formation rate \citep{Padoan95,Krumholz+McKee05sfr,Padoan+Nordlund11sfr,
Hennebelle+Chabrier11sfr,Federrath+Klessen12,Padoan+14PPVI} and of the stellar initial mass function \citep{Padoan+Nordlund97imf,Padoan+Nordlund02imf,
Hennebelle+Chabrier08,Hopkins12imf} are based on the statistics of turbulent fragmentation, such as the probability density function (PDF) of gas density 
and the scalings of velocity and density fluctuations. So far, these statistics have been derived almost exclusively from numerical simulations with rather 
idealized setups, including periodic boundary conditions, driving by a random volume acceleration, isothermal equation of state, or no self-gravity
\citep[e.g.][]{Boldyrev+02,Padoan+04PRL,Kritsuk+07,Federrath+08,Lemaster+Stone08,Price+2011,Federrath13}. These simulations
do not address important issues related to the coupling of the internal dynamics of MCs with the turbulence on larger scales, such as the (large-scale) 
origin of the turbulence and its role in the formation and dispersion of MCs and the cloud's finite lifetime.

To study MC turbulence in a more realistic larger-scale context, and specifically to test the idea that MC turbulence is driven primarily 
by SN explosions, we have carried out a magneto-hydrodynamic (MHD) adaptive-mesh-refinement (AMR) simulation of 
SN-driven turbulence with the Ramses AMR code \citep{Teyssier02}. The numerical method and setup were discussed extensively 
in \cite{Padoan+16mc}--Paper I hereafter--and are only briefly summarized in the next section. Although still somewhat idealized (e.g.~periodic boundary 
conditions and no vertical stratification), this simulation represents a major advance 
relative to previous statistical studies of supersonic turbulence: it allows us i) to test the effect of a realistic and physically motivated driving force, 
such as SN explosions, ii) to see the development of MC turbulence as an integral part of the process of cloud formation and dispersion, iii) to select 
a very large sample of MCs, forming \emph{ab initio} from the large scale turbulence with realistic initial and boundary conditions (and realistic 
statistical distributions of such conditions). 

In Paper I we demonstrated that clouds selected from the simulations
have mass and size distributions, and velocity-size and mass-size relations in agreement with the observations. Using tracer particles, we also
studied their evolution and found that they form and disperse in approximately four dynamical times. We also studied the velocity scaling in the whole
volume and within individual MCs, showing that the turbulence, driven purely by SN explosions, is efficiently injected into MCs, with a realistic velocity
dispersion in the dense gas. In this work, we focus on a specific aspect of direct interest to the modeling of star formation, that is the compressive ratio 
(the ratio of powers in compressive and solenoidal motions) of MC turbulence and its relation to the statistics of density fluctuations 
\citep[e.g.][]{Federrath+08,Schmidt+09,Kritsuk+10,Federrath+10,Kritsuk+11codes}. 

Besides SNe, galactic gas infall, large-scale disk instabilities and spiral arm shocks are also sources of large-scale turbulence 
\citep[e.g.][]{Elmegreen+2003,Bournaud+10,Semenov+15}, but it is generally accepted that SN explosions dominate 
the energy budget of star-forming galaxies at MC scales \citep[e.g.][]{Ostriker+10sfr,Ostriker+Shetty11,Faucher-Giguere+13,Lehnert+13}.
Because the primary goal of this work is to study the compressive ratio of the turbulence in MCs, we focus on the driving by SNe.

Prior to our work, large-scale SN-driven turbulence in the multi-phase ISM has been studied with
fully-periodic volumes without stratifications \citep[e.g.][]{Balsara+04}, or with vertically extended, stratified galactic-fountain simulations 
\citep[e.g.][]{deAvillez07scaling,Joung+09,Hill+12}. These works demonstrated that SNe can drive ISM turbulence with an outer scale of 
$\sim 100$ pc, and derived its velocity scaling laws and gas density PDFs. However, they did not reach the necessary spatial resolution 
to study MC properties, particularly the cascade of SN-driven turbulence in their interior.

More recent simulations in periodic boxes 
\citep{Gatto+15} or stratified galactic-fountains \citep{Walch+15} have significantly lower spatial resolution than earlier works, and do 
not tackle the problem of MC turbulence either. With resimulations of a kpc-size region from a global disc-galaxy simulation, \citet{Dobbs15}
achieved a large enough dynamic range to study the formation and disruption of MCs. In these simulations, MC turbulence
is generated with various prescriptions for SN feedback, resulting in MCs with realistic velocity dispersion. However, the feedback is instantaneously 
inserted in any region of converging flows as the gas density reaches a threshold value of 500 cm$^{-3}$, so the ability of SN feedback to drive
the turbulence within MCs is assumed rather than demonstrated. The statistical properties of SN-driven turbulence from these simulations 
are not discussed.

This paper is structured as follows. In section 2, we give a brief description of the simulation setup, and in section 3 we summarize our recent findings 
concerning the compressibility of 
SN-driven turbulence. Section 4 derives the overall expansion/contraction and rotation of the MCs selected from our simulation. Section 5 analyzes 
the statistics of the compressive ratio of the turbulence within those MCs. The density variance-Mach number relation and the density probability 
distribution in the MCs are explored in sections 6 and 7, respectively. Our conclusions are summarized in section 8.

\section{The Simulation}

We simulate a cubic region of size $L_{\rm box}=250$ pc, with a 
minimum cell size of $dx=0.24$ pc (a maximum resolution equivalent to a mesh of $1024^3$ cells), periodic boundary conditions,
a mean density of 5 cm$^{-3}$ (corresponding to a total mass of $1.9\times 10^6$ M$_{\odot}$) and a core-collapse SN rate of 6.25 Myr$^{-1}$. 
We distribute SN explosions randomly in space and time (see discussion in Paper I in support of this choice), so our SN 
rate could also be interpreted as the sum of all types of SN explosions. Individual SN explosions are implemented with an instantaneous 
addition of 10$^{51}$ erg of thermal energy and 15 M$_{\odot}$ of gas, distributed with an exponential profile in a spherical 
region of radius $r_{\rm SN}=3 dx=0.73$ pc, which guarantees numerical convergence of the SN remnant evolution \citep{Kim+Ostriker15SN}. 

Besides the pdV work, and the thermal energy introduced to model SN explosions, our total energy equation adopts 
uniform photoelectric heating up to a critical density of 200 cm$^{-3}$, and parametrized cooling functions from \citet{Gnedin+Hollon12}.
The simulation is started with zero velocity, a uniform density $n_{\rm H,0}=5$ cm$^{-3}$, a uniform magnetic field $B_0=4.6$ $\mu$G and a 
uniform temperature $T_0=10^4$ K. The first few SN explosions rapidly bring the mean thermal, magnetic and kinetic energy to approximately 
steady-state values, with the magnetic field amplified to an rms value of 7.2 $\mu$G. We have run the simulation for 45 Myr without self-gravity 
and then continued with self-gravity for 11 Myr. The interested reader is referred to Paper I for further details about the numerical setup.

\section{Compressibility of SN-Driven Turbulence}

Before analyzing the compressive ratio of the turbulence within MCs, we briefly summarize our recent results on the overall compressibility of SN-driven 
turbulence (see details in Paper I). In our discussion, we shall make a strict distinction between the driving acceleration, ${\bs a}$, for the 
turbulent velocity and the driving force, $F\equiv \rho {\bs a}$, for the flow momentum. As in previous works on supersonic turbulence, we are more concerned 
with the effective driving acceleration, ${\bs a}$, rather than the driving force, $F$, because the compressibility of the velocity field is directly related to that of 
the driving acceleration, not the force. All studies of the compressibility of interstellar turbulence decompose the velocity, ${\bs v}$, rather than the momentum, 
$\rho{\bs v}$, into solenoidal and compressive modes, and refer to the compressibility of the driving acceleration, rather than that of the driving force 
\citep[e.g.][]{Schmidt+09,Kritsuk+10,Federrath+10,Kritsuk+11codes,Federrath13}.

In our simulation, the SN explosion energy is deposited as thermal energy in small, randomly-selected spheres, 
so it is initially injected via the pressure term in the Navier-Stokes equation. 
Denoting as $P_{\rm s}$ the pressure source due to SN explosions, the effective driving force and 
acceleration can be written as  $-\nabla P_{\rm s}$ and $-(\nabla P_{\rm s})/\rho$, respectively. 
Although  the effective force, $-\nabla P_{\rm s}$, is purely compressive,  the driving acceleration, $-(\nabla P_{\rm s})/\rho$, 
is not so, in general.
Clearly, the divergence and the curl of this effective acceleration, $-(\nabla P_{\rm s})/\rho$, are given, respectively, 
by $(\nabla P_{\rm s} \cdot \nabla \rho)/\rho^2 - (\nabla^2 P_{\rm s})/\rho$ and $(\nabla P_{\rm s} \times 
\nabla \rho)/\rho^2$. The latter, known as the baroclinic effect \citep[e.g.][]{Passot+Pouquet87,Vazquez-Semadeni+96}, is nonzero in general. 
In particular, considering random density and pressure fluctuations outside a SN-explosion sphere, 
the baroclinic term is always nonzero at the boundary of such a sphere.  Therefore, the effective acceleration for SN driving is neither 
purely compressive nor purely solenoidal, rather it consists of a mixture of solenoidal and compressive modes.  
If the direction of  $\nabla P_{\rm s}$ is random with respect to $\nabla \rho$, the divergence 
and curl of $-(\nabla P_{\rm s})/\rho$ are comparable, suggesting similar amounts 
of solenoidal and compressive modes in the effective acceleration. 
With the expansion of the remnant, solenoidal motions generated around the pressure ``sources" 
are transferred to larger scales. For example, a single SN remnant leads 
to an energy spectrum that peaks at a wavenumber $k \simeq 1/R$, with $R$ the remnant 
radius (see Paper I). We find that, at large scales, the solenoidal and compressive 
modes in our simulation are roughly in equipartition, 
supporting the above picture.  

\begin{figure}[t]
\includegraphics[width=\columnwidth]{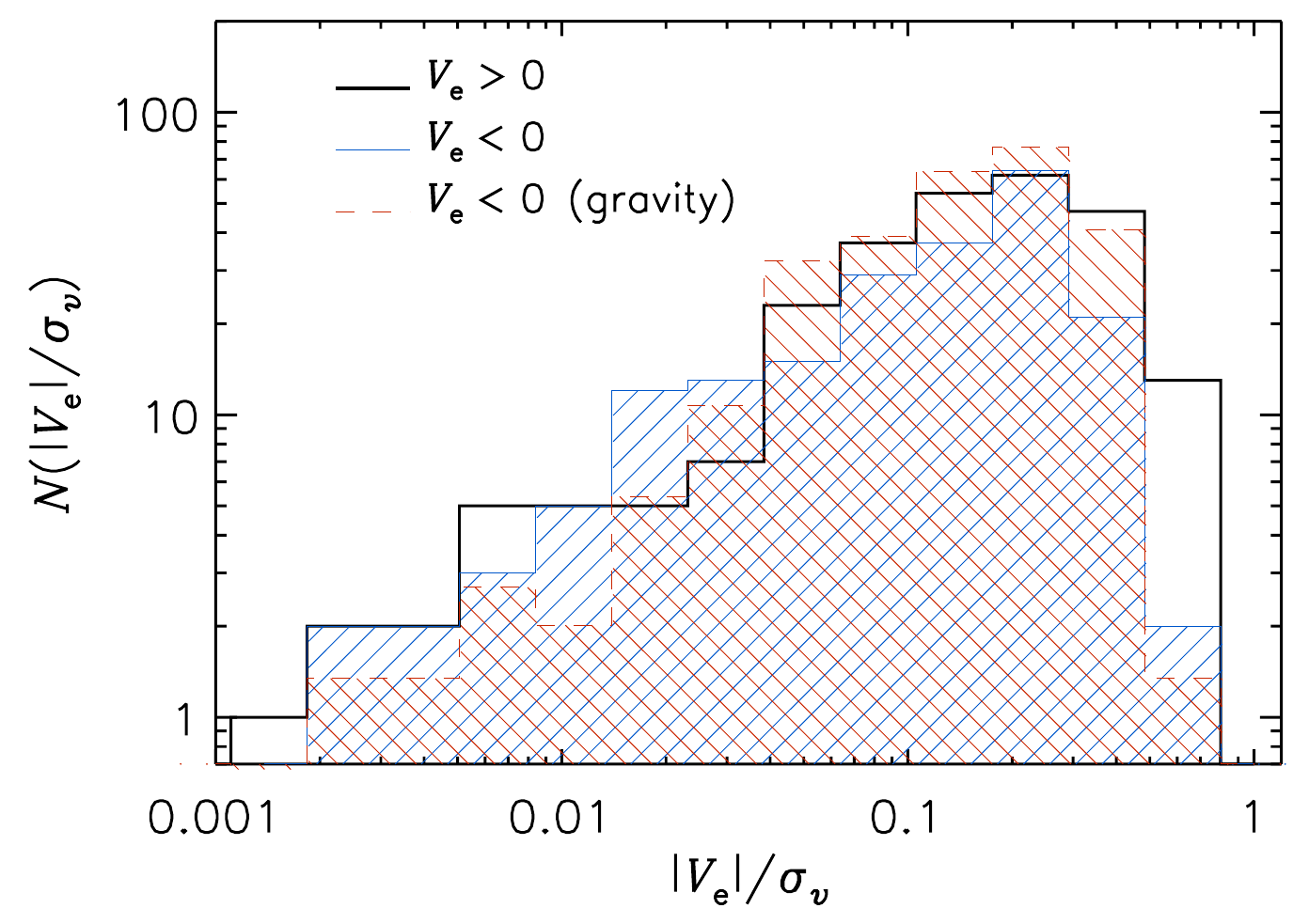}
\caption[]{Probability distribution of mean cloud expansion velocity, $V_{\rm e}>0$ (unshaded histogram), and contraction 
velocity, $V_{\rm e}<0$ (solid-line shaded histogram), normalized to the rms velocity in the cloud, for a sample of 507 clouds 
selected before the introduction of gravity. The dashed-line shaded histogram shows the probability distribution of mean cloud 
contraction velocity for a sample of 802 clouds selected after gravity is included in the simulation. This histogram has been 
normalized to the same total probability as the other two histograms.}
\label{vex_hist}
\end{figure}

While the remnant expansion brings the velocity power to large scales, the 
nonlinear advection term causes cascades of both solenoidal and compressive modes 
towards small scales. If the SN rate is not too high and there is sufficient time in 
between SN events to allow the flow to fully develop, a dynamically quasi-relaxed state 
is reached. During the relaxation phase, one might expect the interaction 
between solenoidal and compressive modes via the nonlinear term to establish an 
equipartition between the two modes in the inertial range, as seen in simulations adopting an 
isothermal equation of state and purely solenoidal driving. However, this inertial-range equipartition is not observed in our simulation, 
because, in the relaxation phase, the baroclinic effect preferentially converts compressive modes
(shocks or expansions) to solenoidal motions. Due to its dependence on density and pressure gradients, the baroclinic effect is 
more efficient at smaller scales, and, as a result, the compressive spectrum decreases towards small 
scales faster than the solenoidal one (see Figure 8 in Paper I).  As a consequence, at scales corresponding to 
MC sizes, the ratio of compressive to solenoidal power is, on average, below the equipartition value of 0.5. 
The goal of this work is to derive the distribution of the compressive ratio of SN-driven turbulence in MCs 
and the corresponding amplitude and probability distribution of density fluctuations.

\section{Overall Expansion/Contraction and Rotation of Molecular Clouds}

MCs may have non-negligible mean radial motions (expansion or contraction) and overall rotation. Although these mean motions 
can be considered as the large-scale components of MC turbulence, it is nonetheless interesting to examine their importance relative to the random 
motions. In the next section, we will consider the compressive ratio of MC turbulence both with and without the contribution of the mean radial motion 
and of the overall rotation.

\begin{figure}[t]
\includegraphics[width=\columnwidth]{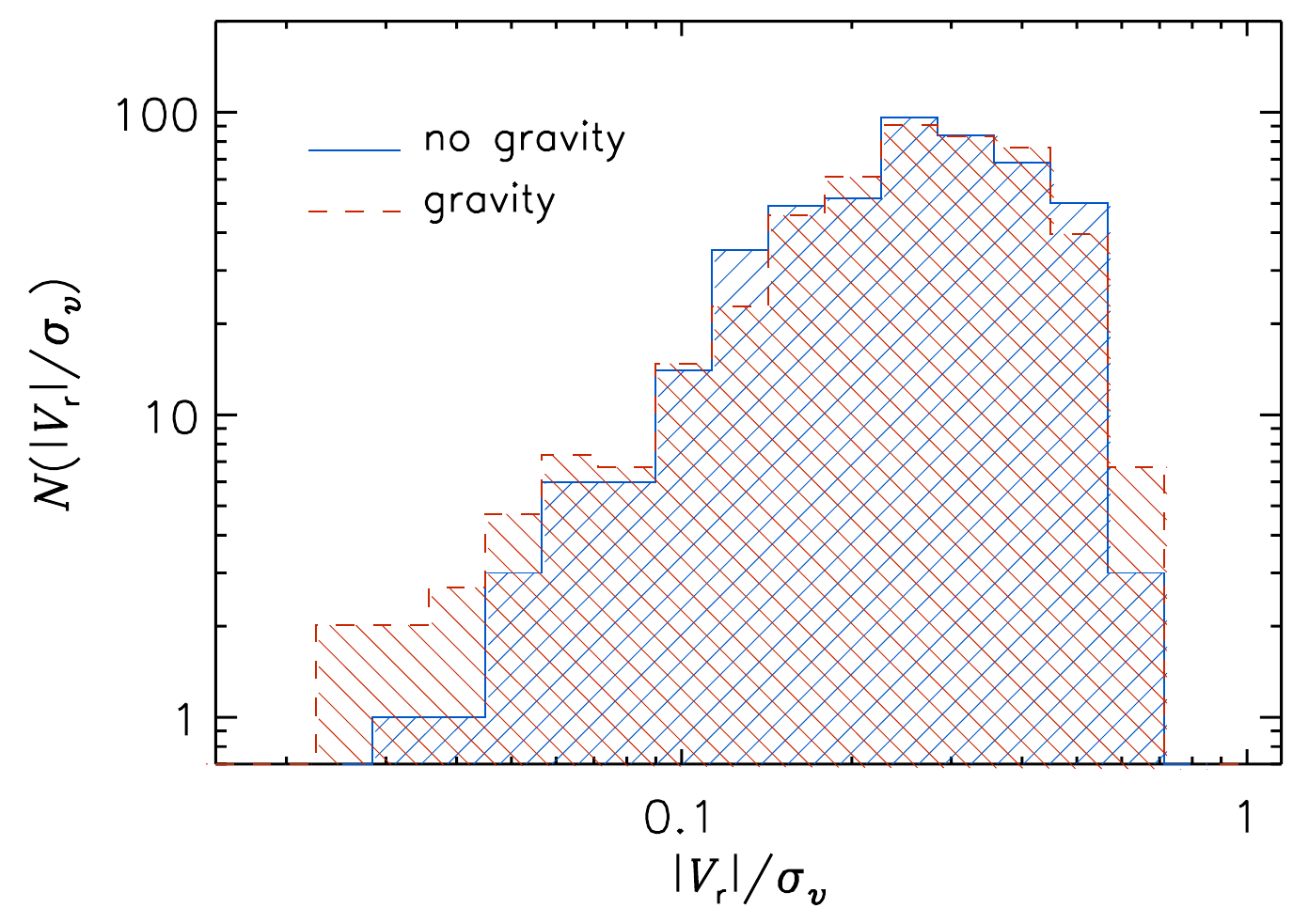}
\caption[]{Probability distribution of the cloud solid-body rotational velocity, $V_{\rm r}$, for the same cloud samples as in Figure \ref{vex_hist}, 
selected before and after the inclusion of gravity.}
\label{vrot_hist}
\end{figure}

We select MCs in the simulation as connected regions above a density threshold of $n=200$ cm$^{-3}$ (see Paper I) and with 
mass $M_{\rm cl}> 1,000$ M$_{\odot}$, and analyze the velocity field in the volume within the smallest rectangular cuboid (`bounding box' hereafter) 
containing each cloud. The actual cloud may cover only a small fraction of the total volume of its bounding box, so the velocity field we
analyze includes lower-density, possibly warmer, gas. The inclusion of this surrounding gas in the analysis of the velocity field is justified
because the dynamical evolution of the cloud involves the surrounding region, with lower density gas accreting onto the cloud, and 
denser gas expanding out of the cloud. The cloud boundary at 200 cm$^{-3}$ only serves the purpose of selecting individual objects and
has no special dynamical significance.  

The overall expansion rate of a cloud is evaluated as $\frac{1}{3}\langle \nabla \cdot {\bs v}\rangle$, 
with $\langle \nabla \cdot {\bs v}\rangle$ the mean velocity divergence in the cloud bounding box. Assuming a 
uniform expansion velocity, $\frac{1}{3} \langle \nabla \cdot {\bs v}\rangle {\bs r}$, with 
${\bs r}$ the separation to the cloud center, we define a characteristic expansion velocity, $V_{\rm e}$, as
\begin{equation}
V_{\rm e} \equiv \left ( \frac{1}{9 V_{\rm bb}} \langle \nabla \cdot {\bs v}\rangle ^2\int_{V_{bb}} {\bs r}^2 d{\bs r} \right)^{1/2} =  \frac{1}{3} |\langle \nabla \cdot {\bs v}\rangle| R_{\rm c},
\label{eq_vex}
\end{equation}
where $V_{\rm bb}$ is the volume of the bounding box and $R_{\rm c}$ is the effective cloud radius,
\begin{equation}
R_{\rm c} = \frac{1}{2}[(L_{\rm x}^2 + L_{\rm y}^2 +L_{\rm z}^2)/3]^{1/2},
\label{eq_rc}
\end{equation}
with $L_{\rm x, y, z}$ the sides of the minimum bounding box. The absolute value, $|V_{\rm e}|$, is the rms of the 
overall expansion velocity. We find a characteristic value of $|V_{\rm e}|\sim 1$ km s$^{-1}$. 
To evaluate the dynamical importance of $V_{\rm e}$, we plot in Figure \ref{vex_hist} the 
histograms of $|V_{\rm e}|$ divided by the total rms velocity, $\sigma_{\rm v}$, for 507 clouds 
selected from 30 snapshots before the introduction of gravity in the simulation, and 802 clouds selected from 30 snapshots after gravity was included. 
The histograms show that $V_{\rm e}$ is typically only 10-20\% of $\sigma_{\rm v}$.  
%meaning that, on the average, more than 95\% of the internal kinetic energy is in the 
%form of random motions. 
For expanding ($V_{\rm e}>0$) clouds, $\langle V_{\rm e}/\sigma_{\rm v}\rangle = 0.15$ and 0.19, 
in the cases with and without gravity respectively, while for contracting ($V_{\rm e}<0$) clouds, 
$\langle |V_{\rm e}|/\sigma_{\rm v}\rangle = 0.15$ in both cases. The comparison of the mean values for contracting clouds with and without gravity and their very similar
histograms shown in Figure \ref{vex_hist} demonstrate that, despite the presence of 
self-gravity causing the collapse of their dense cores, MCs do not undergo global collapse. 
However, the fraction of clouds that are contracting (rather than expanding) is 0.62 with gravity and 0.45 without gravity, suggesting that gravity
may be causing a global contraction in at least a fraction of the clouds (even if their kinetic energy is dominated by random motions).

\begin{figure}[t]
\includegraphics[width=\columnwidth]{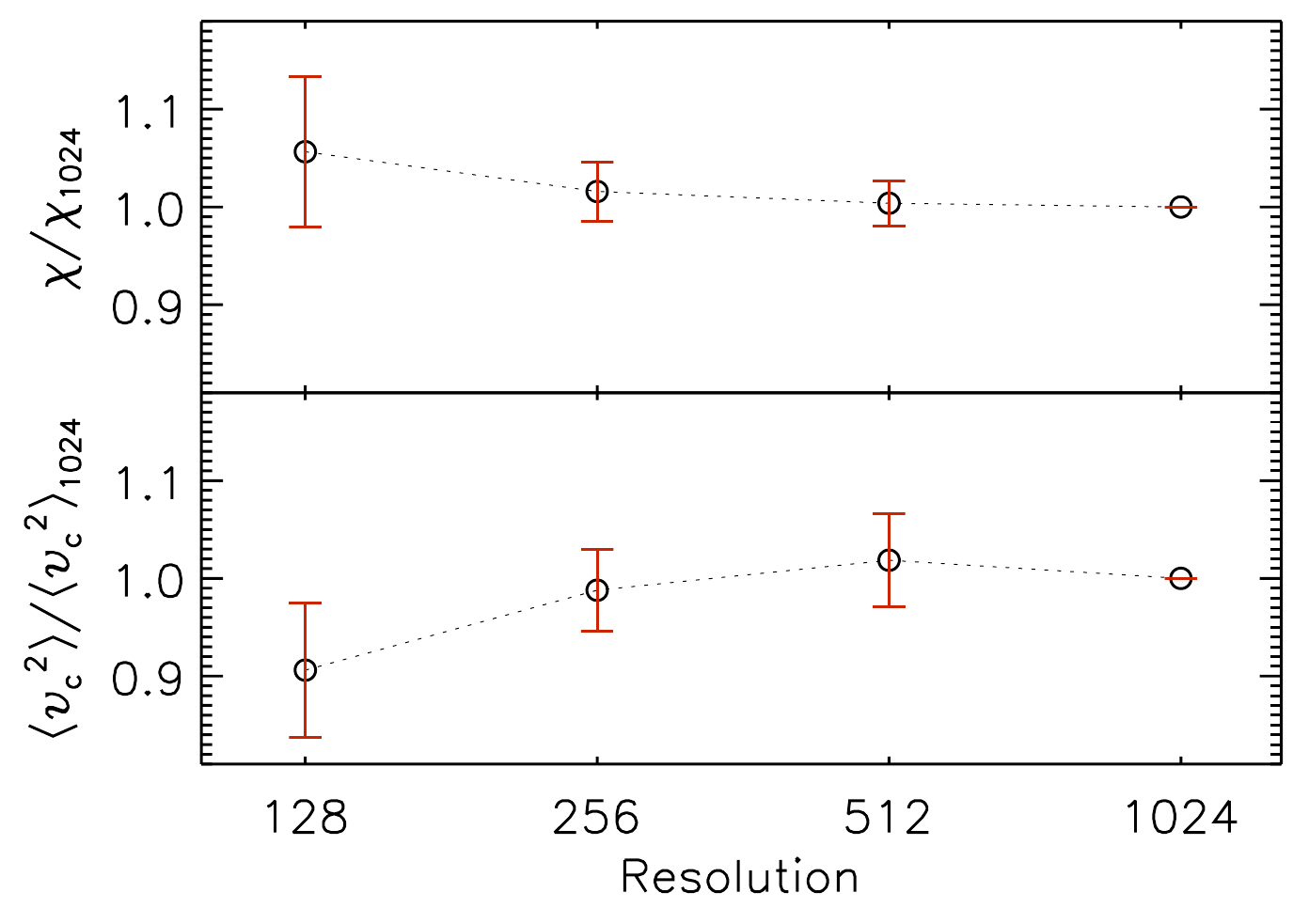}
\caption[]{Convergence test of the measured cloud compressive ratio, $\chi$, and compressive power, $\langle v_{\rm c}^2 \rangle$, normalized
to their values when the clouds are extracted at the maximum resolution, $\chi_{1024}$ and $\langle v_{\rm c}^2 \rangle_{1024}$}
\label{converge}
\end{figure}

The rate of the overall rotation of a cloud can be characterized by an angular velocity 
${\bs \Omega} = \frac{1}{2}\langle {\bs \omega} \rangle$, with $\langle {\bs \omega} \rangle$ the average vorticity in the 
cloud. The velocity field of a solid-body rotation is given by  ${\bs \Omega} \times {\bs r}$, so we define the rms velocity of 
the overall rotation in the bounding box of each cloud, $V_{\rm r}$, as 
\begin{equation}
V_{\rm r}^2 = \frac{1}{4 V_{\rm bb}} \int_{V_{bb}} (\langle {\bs \omega} \rangle \times {\bs r} )^2 d {\bs r}.
\label{eq_vrot}
\end{equation}
Figure \ref{vrot_hist} shows that the histograms of $V_{\rm r}/ \sigma_{\rm v}$ of clouds before (solid line) and after (dashed line) 
the inclusion of self-gravity are quite similar, with mean values of 0.28 and 0.27, respectively. The comparison with Figure \ref{vex_hist}
also illustrates that the solid-body rotation of the clouds contains more energy than the overall expansion or contraction, perhaps because the rotation 
has a larger number of degrees of freedom than the mean radial motion.

\section{Compressive Ratio in Molecular Clouds}

We define the compressive ratio, $\chi$, of the velocity field, ${\bs v}$, in a MC as
\begin{equation}
\chi \equiv \langle {\bs v}_{\rm c}^2 \rangle/  \langle {\bs v}_{\rm s}^2 \rangle, 
\label{eq_chi}
\end{equation}
where ${\bs v}_{\rm c}$ and ${\bs v}_{\rm s}$ are the compressive and solenoidal components of ${\bs v}$ in the cloud bounding box. 
The two velocity components are derived with the standard Helmholtz decomposition
in Fourier space. We have verified that the values of $\chi$ are quite insensitive to the boundary of the bounding box: Making the velocity field in
the bounding box periodic with a tapered cosine window function (a gradual drop to zero of the
velocity, affecting only the three outermost cell layers) only changes the measured $\chi$ within a few percent.

As an independent check, we also measured $\chi$ based on the longitudinal ($S_{\rm LL}$) and transverse ($S_{\rm NN}$) 
structure functions in each cloud. Under the 
assumption of statistical isotropy, exact relations exist between the structure functions and the compressive and solenoidal power spectra 
($E_{\rm c}$, $E_{\rm s}$). Using equations (12.35) of \citet{Monin+Yaglom75}, 
we find that $  \langle {\bs v}_{\rm c}^2 \rangle=2 \int E_{\rm c} (k) dk = u^{\prime^2} + \int_0^{\infty}[S_{\rm LL} (r) - S_{\rm NN} (r)] /r \,dr$ 
and $\langle {\bs v}_{\rm s}^2 \rangle =2 \int E_{\rm s}(k) dk = 2u^{\prime^2} - \int_0^{\infty}[S_{\rm LL} (r) - S_{\rm NN} (r)]/r \,dr$, 
where $u^{\prime}$ is the 1D velocity dispersion. The value of $\chi$ computed from this method 
shows a tight correlation with that from the Helmholtz decomposition, confirming the reliability of the measurement. Below, we will 
only consider results from the Helmholtz decomposition.

Our derived values of $\chi$ are expected to be numerically converged, because the selected clouds are very well resolved 
(see images of selected clouds in Figure 3 of Paper I), with bounding box volumes between $26^3$ and $340^3$ computational cells 
(only 3\% of the boxes are below $40^3$ cells) and because $\chi$ is a large-scale quantity  within each cloud 
bounding box, as it only depends on total powers. To illustrate the large-scale nature of $\chi$ and the fact that it is insensitive to spatial resolution, 
we have analyzed the selected clouds at four different resolutions (of the same simulation), corresponding to a range of $128^3$ to $1024^3$ cells in the whole 
computational volume.  Figure \ref{converge} shows this convergence test, where $\chi$ and $\langle v_{\rm c}^2 \rangle$ have been normalized to 
their values at the highest resolution and then averaged over all clouds. Convergence is clearly achieved at $512^3$, and deviations in $\chi$
are within the 1-$\sigma$ uncertainty even at the lowest resolution.
 
\begin{figure}[t]
\includegraphics[width=\columnwidth]{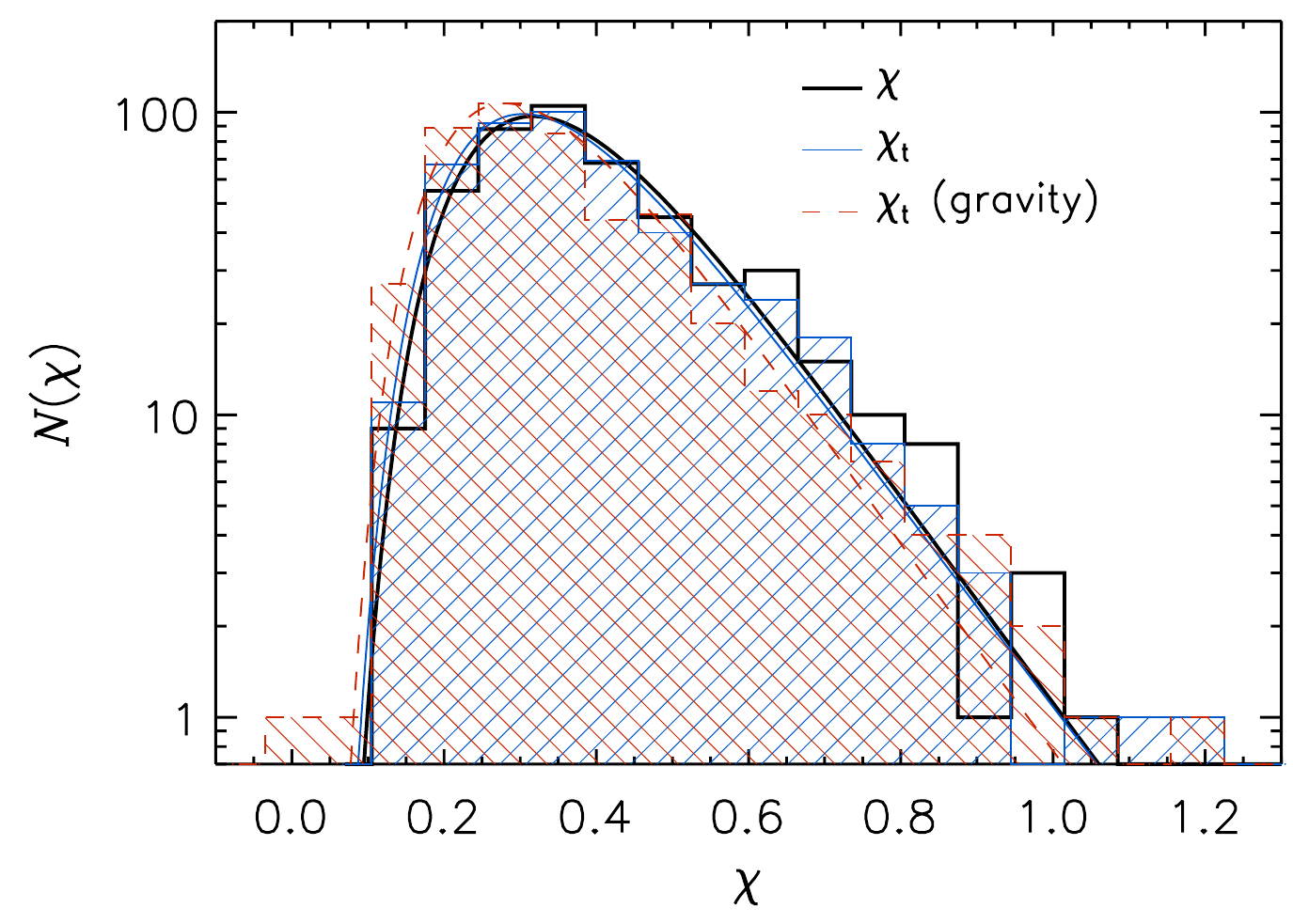}
\caption[]{Probability distributions of the total compressive ratio, $\chi$, (unshaded, thick solid line histogram) and its turbulent component, 
$\chi_{\rm t}$ (shaded histograms). The probability of $\chi_{\rm t}$ is plotted for clouds selected before (solid-line histogram) 
and after (dashed-line histogram) the inclusion of gravity (the same cloud samples as in Figures \ref{vex_hist} and \ref{vrot_hist}). 
The smooth curves are lognormal fits with mean values of  -1.16, -1.19 and -1.27 and rms values of 0.38, 0.40 and 0.41 for $\chi$, $\chi_{\rm t}$ 
without gravity and $\chi_{\rm t}$ with gravity respectively.}
\label{chi_hist}
\end{figure}

The solid unshaded histogram in Figure \ref{chi_hist} shows the distribution of $\chi$ for the 507 MCs selected before the inclusion of self-gravity. 
The histogram is well approximated by a lognormal distribution (with the mean and rms of $\ln (\chi)$ equal to -1.16 and 0.38, respectively) that 
peaks at $\chi \simeq 0.31$, below the equipartition value of 0.5 found in isothermal simulations of highly supersonic turbulence with purely solenoidal 
driving \citep[e.g.][]{Kritsuk+10,Kritsuk+11codes,Federrath13} and the value of $\approx 1.0$ from isothermal simulations with purely compressive driving 
\citep[e.g.][]{Federrath+10,Federrath13}, and comparable to the value found in isothermal simulations with random solenoidal driving and with a relatively low
sonic or Alfv\'{e}nic Mach numbers $\approx 3$ \citep[][]{Kritsuk+10,Kritsuk+11codes}. As discussed in section 3 and in Paper I, this low mean 
value of $\chi$ at MC scales is likely the result of the baroclinic effect (absent in isothermal simulations) making the power spectrum of the solenoidal modes 
much shallower than that of the compressive modes (see further discussion in section 8). 

The overall expansion (or contraction) of a cloud contributes to the power in compressive motions, $\langle {\bs v}_{\rm c}^2 \rangle$, 
while the solid-body rotation contributes to the solenoidal power, $\langle {\bs v}_{\rm s}^2 \rangle$; the power ratio of compressive to 
solenoidal motions may be computed after subtracting these contributions. Thus, for each cloud, we define a new turbulent compressive ratio,
$\chi_{\rm t}$, as 
\begin{equation}
\chi_{\rm t} \equiv [\langle {\bs v}_{\rm c}^2 \rangle- V_{\rm e}^2]/[\langle {\bs v}_{\rm s}^2 \rangle - V_{\rm r}^2].
\label{chit}
\end{equation}
The shaded solid-line histogram in Figure \ref{chi_hist} shows the probability distribution of $\chi_{\rm t}$ 
for the same clouds as in the unshaded histogram of $\chi$.  It turns out that, after subtracting the contributions of the overall 
expansion and rotation, the distribution of the compressive ratio remains largely unchanged. The PDF of $\chi_{\rm t}$ also peaks 
around 0.3, confirming that the majority of {\it turbulent} energy is in solenoidal motions.    
%with the mean value decreasing from $\langle \chi \rangle=0.42$ to $\langle \chi_{\rm t} \rangle=0.33$ ({\it Needs to be revised here when we have final results}).
The shaded dashed-line histogram in Figure \ref{chi_hist} shows the distribution of $\chi_{\rm t}$ for the clouds selected after the inclusion of self-gravity. 
Interestingly, the histograms of $\chi_{\rm t}$ for the cases with and without gravity appear to be very similar as well (peaking at $\chi_{\rm t} = 0.28$ and 0.30, 
respectively), demonstrating that the presence of gravity does not noticeably affect the compressive ratio of turbulent motions in MCs. 

\begin{figure}[t]
\includegraphics[width=\columnwidth]{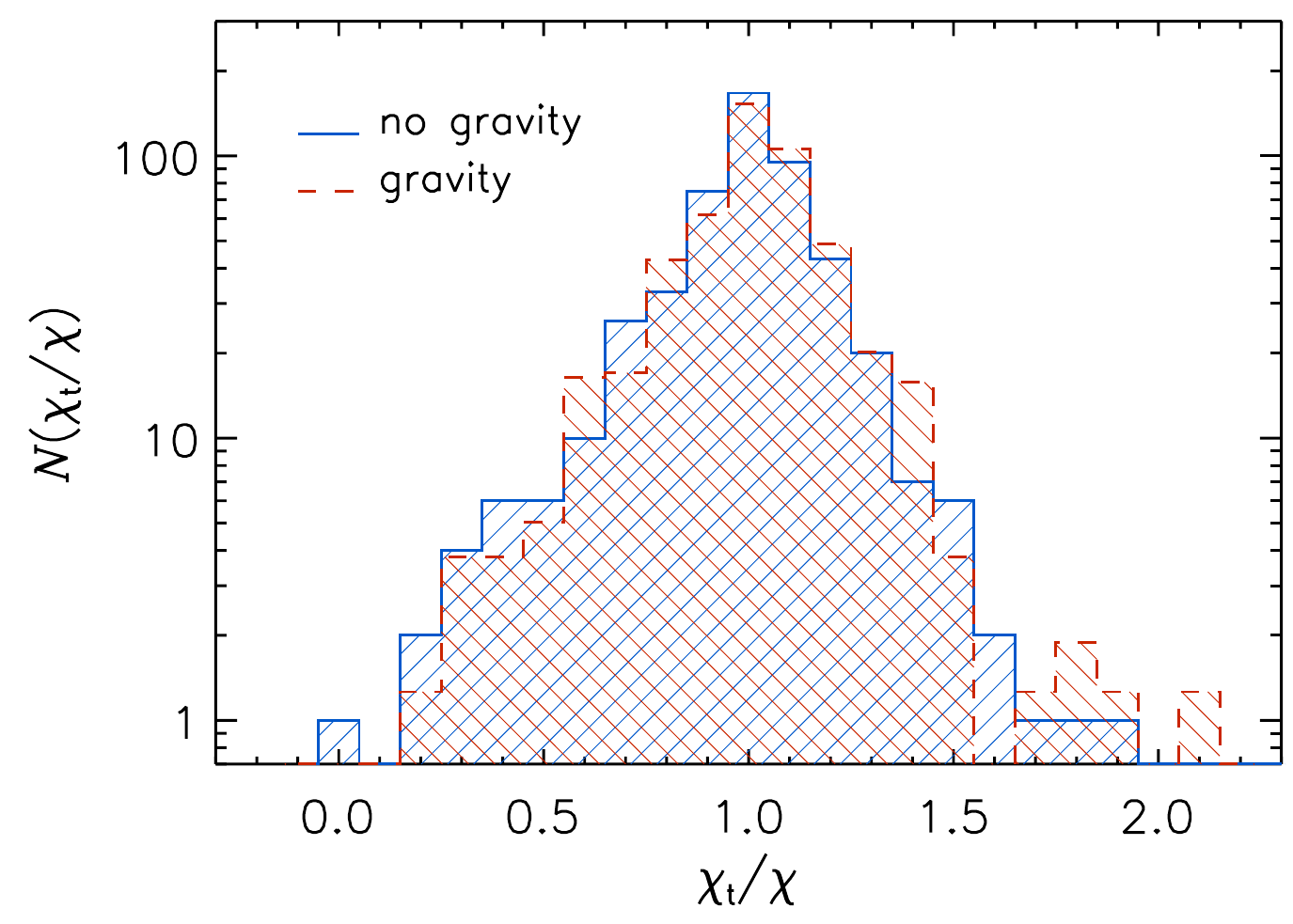}
\caption[]{Probability distribution of $\chi_{\rm t}/\chi$ for the same cloud samples with and without gravity as in the previous figures.}
\label{h_chi}
\end{figure}

Although the probability distributions of $\chi$ and $\chi_{\rm t}$ shown in Figure \ref{chi_hist} are nearly identical, the subtraction of the 
mean radial and rotational motions from the compressive ratio can have quite a large effect for individual clouds. Figure \ref{h_chi} shows
the probability distribution of the ratio $\chi_{\rm t}/\chi$ for the same samples of clouds with and without gravity as in the previous figures. 
Although centered around a mean value $\approx 1.0$, the distribution is quite broad. It is also rather insensitive to gravity.

%
%\begin{figure}[t]
%\includegraphics[width=\columnwidth]{chi_t_over_chi.pdf}
%\caption[]{}
%\label{chi}
%\end{figure}
%

%

\section{Density Fluctuations versus Mach Number}

Using isothermal simulations, it has been shown that the rms density, $\sigma_{\rho/\rho_0}$, where $\rho_0$ is the mean density, 
scales linearly with the rms Mach number of the flow, $\cal{M}$, or, introducing the logarithm of the density, $s\equiv\ln (\rho/\rho_0)$, 
\begin{equation}
\sigma_s^2 = \ln (1 + b^2 {\cal M}^2),  
\label{rms}
\end{equation}
where $b\approx 1.0$ if the turbulence is driven by a purely compressive acceleration, $b\approx 0.3$ if the acceleration is solenoidal, and even smaller values 
are possible with magnetic fields \citep[e.g.][]{Padoan+97ext,Nordlund+Padoan99pdf,Federrath+10,Padoan+Nordlund11sfr,Price+2011,Molina+12}. 
Expressions different from (\ref{rms}) \citep[e.g.][]{Ostriker+01,Lemaster+Stone08} or extensions to non-isothermal polytropic or adiabatic turbulence 
\citep{Nolan+15,Federrath+Banerjee15} have also been proposed. Because $\sigma_{\rm s}$ is a crucial quantity in models of star formation based 
on turbulent fragmentation, it is important to verify if equation (\ref{rms}) holds with realistic energy equation and SN driving as well.

\begin{figure}[t]
\includegraphics[width=\columnwidth]{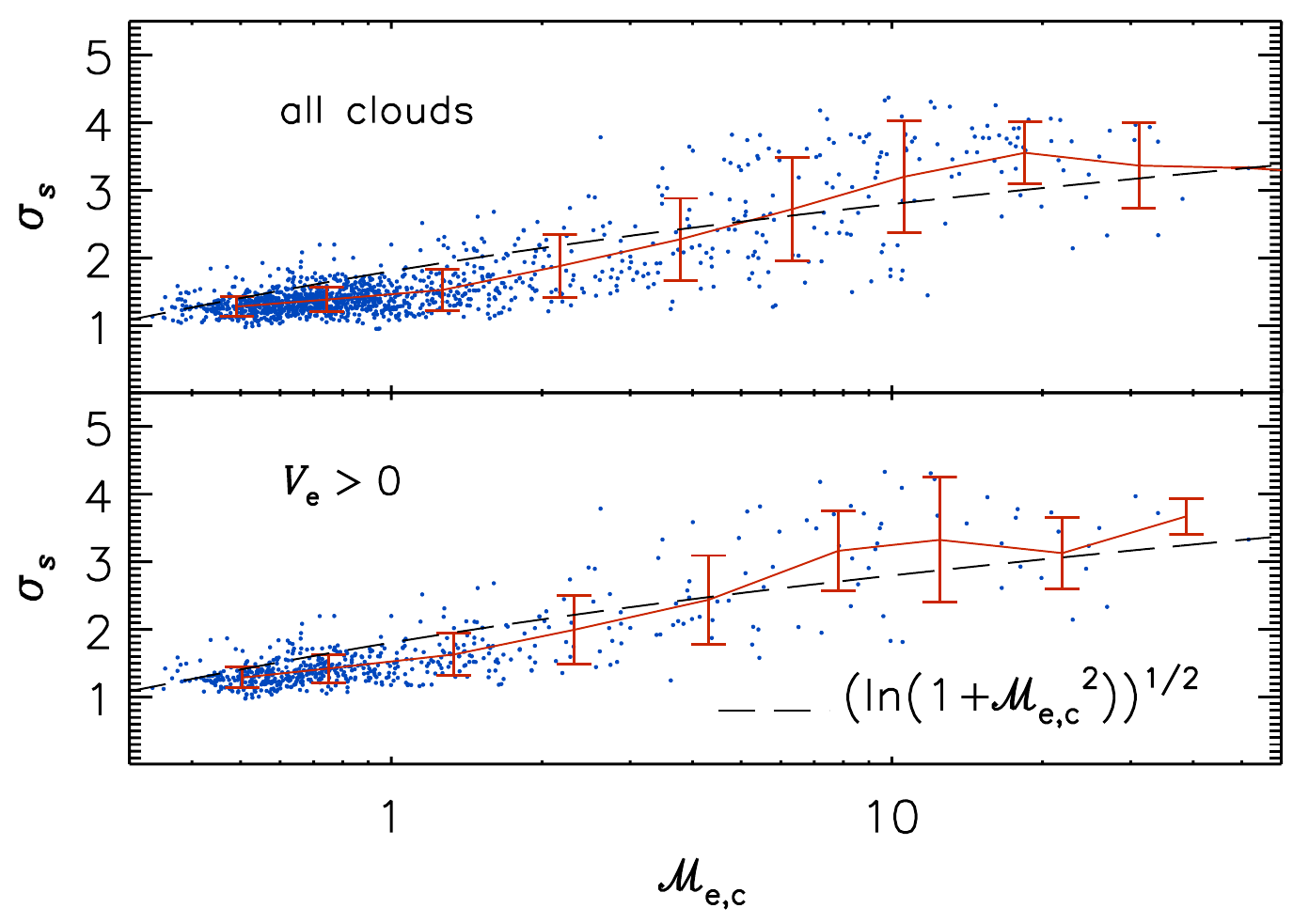}
\caption[]{Rms of logarithmic density versus rms effective Mach number in all clouds (upper panel), and in clouds with positive mean velocity divergence
(lower panel). The solid line and error bars show the mean and standard deviation of $\sigma_s$ computed in logarithmic intervals of ${\cal M}_{\rm e,c}$.
The long-dashed line is the model prediction.}
\label{mach}
\end{figure}

Considering the strong correlation between the parameter $b$ and the compressive ratio \citep{Federrath+10,Konstandin+12}, and because only the 
compressive part of the velocity field can cause density fluctuations, we make the ansazt that, in eq. (\ref{rms}), $b{\cal M}$ is simply the compressive 
component of the Mach number, $b {\cal M}=\sqrt{\chi/(1+\chi)}{\cal M}\equiv {\cal M}_{\rm c}$. Furthermore, to account for the effect of magnetic pressure,
we adopt the model in \citet[][eq. (28)]{Padoan+Nordlund11sfr} and obtain:
\begin{equation}
\sigma_s^2 = \ln (1 + (\beta/(1+\beta))\,{\cal M}_{\rm c}^2 )=\ln (1 + {\cal M}_{\rm e,c}^2 ),  
\label{rms2}
\end{equation}
where $\beta$ is the ratio of gas to magnetic pressure (see also \citet{Molina+12}) and ${\cal M}_{\rm e,c}$ is a compressive effective Mach number 
that includes the effect of magnetic pressure.  

To test this relation, we have computed $\beta$ from the ratio of the Alfv\'{e}nic and sonic rms Mach numbers, $\beta=2 ({\cal M}_{\rm A}/{\cal M})^2$,
where ${\cal M}=\langle(v/c_{\rm s})^2\rangle^{1/2}$, and ${\cal M}_{\rm A}=\langle(v/v_{\rm A})^2\rangle^{1/2}$, with $c_{\rm s}$ and $v_{\rm A}$ the local sound speed
and Alfv\'{e}n velocity respectively. Similarly, we have computed the compressible rms Mach number as ${\cal M}_{\rm c}=\langle(v_{\rm c}/c_{\rm s})^2\rangle^{1/2}$,
where $v_{\rm c}$ is the modulus of the compressive part of the local velocity. Figure \ref{mach} shows $\sigma_{\rm s}$ versus ${\cal M}_{\rm e,c}$ for the 
MCs in our simulation. The upper panel shows the full sample, while the lower one includes only clouds with positive mean velocity divergence ($V_{\rm e}>0$).
As shown by the mean values of $\sigma_s$ computed in logarithmic intervals of ${\cal M}_{\rm e,c}$, the density fluctuations in the MCs from the simulation
are roughly consistent with eq. (\ref{rms2}), particularly in the case of expanding clouds, probably because expanding MCs are, on average, older than contracting 
ones and  thus more relaxed. However, the relation has a very large scatter, so one should not expect a precise correlation between rms density and rms
Mach number in real MCs \citep[e.g.][]{Price+2011}.

\section{Density PDF and Star Formation}

\begin{figure}[t]
\includegraphics[width=\columnwidth]{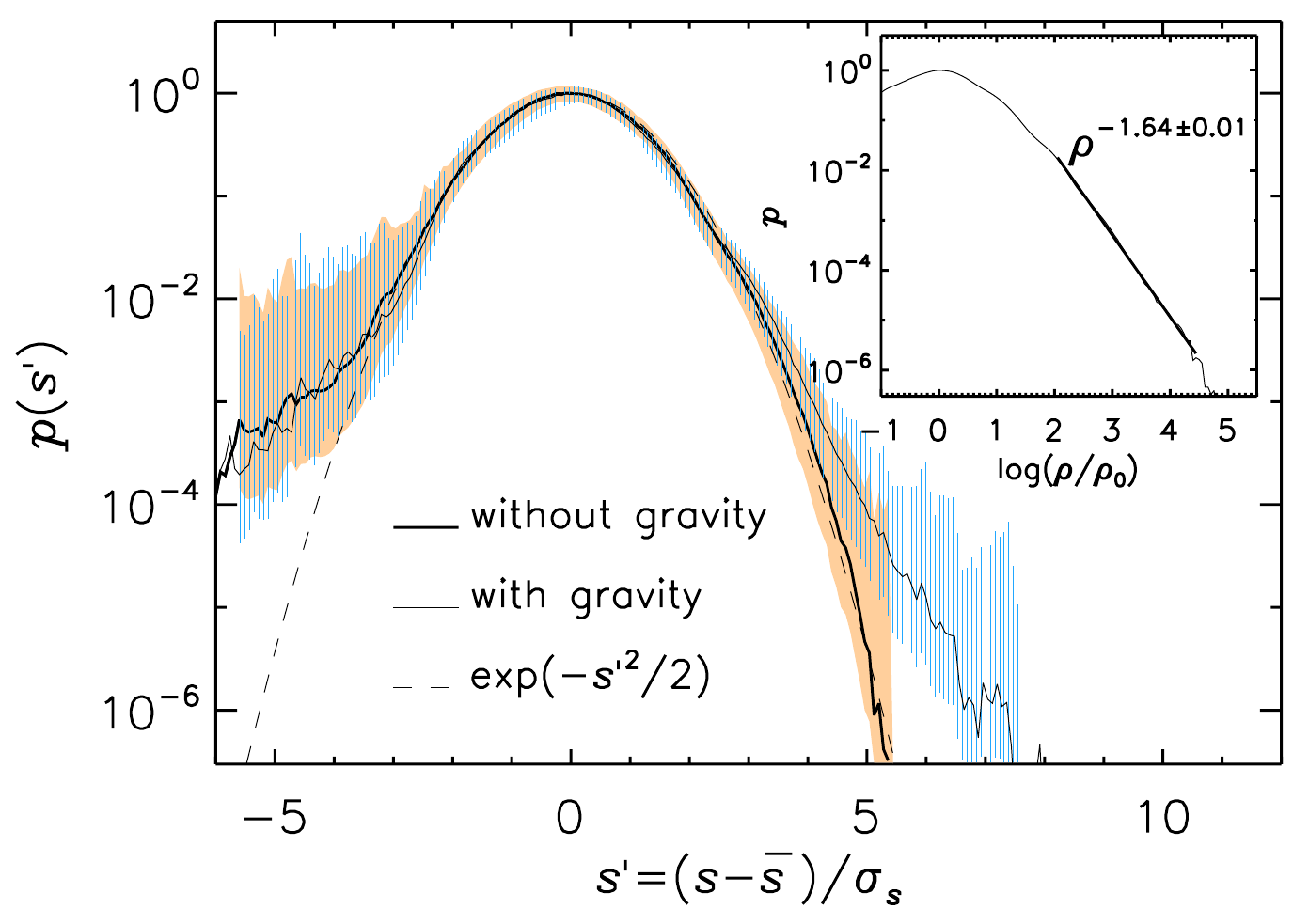}
\caption[]{Composite PDF of logarithmic gas density for clouds selected before (thick line) and after (thin line) including 
gravity. Before averaging the PDFs together, each of them is normalized to zero mean and unity 
rms and divided by its maximum probability value, so the composite PDF represents the average PDF shape. 
The shaded regions show the cloud-to-cloud rms variations. 
The dashed line is the Gaussian fit. Notice that $\bar{s}$ and $\sigma_{\rm s}$
vary from cloud to cloud. {\it Inset}: Sum of all PDFs without any shifting or normalization, for clouds with gravity. 
The slope of the PDF tail is fit by a power law with exponent $-1.64\pm0.01$, in the range $2 < \log(\rho/\rho_0) < 4.5$, corresponding to a slope of -2.64
for the PDF of $\rho$. Notice that the mean density, $\rho_0$, varies from cloud to cloud.}
\label{pdf}
\end{figure}

The compressive ratio of the turbulence has been shown to affect also the probability density function 
(PDF) of density fluctuations. The PDF is nearly lognormal in isothermal supersonic flows 
driven by a prescribed large-scale solenoidal acceleration \citep[e.g][]{Kritsuk+07}, though different function 
forms have also been proposed \citep[e.g.][]{Hopkins+13}. However, if the driving acceleration is purely compressive, 
the PDF exhibits significant negative skewness \citep[e.g.][]{Federrath+10}. Here we analyze the distribution 
of $s$ ($\equiv \ln(\rho/{\rho_0})$) in the MCs selected from our simulation, where the effective 
SN driving consists of a mixture of compressive and solenoidal modes.

To focus on the PDF shape, we normalize the measured PDF of $s$ in the bounding 
box of each cloud, $p_s$, to a mean of zero and an rms of one. The normalized PDF 
corresponds to the distribution of $(s-\bar{s})/\sigma_{s}$, which would be Gaussian with 
zero mean and unity rms if the density PDF were exactly lognormal. The thick solid line 
in Figure \ref{pdf} shows the average of the normalized PDFs in the clouds selected 
before the inclusion of gravity. This composite PDF is in agreement  with a Gaussian distribution (dashed line), 
except for an excess of probability on the left tail. The excessive left tail causes a small negative 
skewness, $-0.13\pm 0.03$ (where $\pm 0.03$ corresponds to the uncertainty in the measurement of the mean skewness).
This value is consistent with that found in simulations with solenoidal driving, $-0.10\pm 0.01$, and significantly smaller than that from purely compressive driving, 
$-0.26\pm 0.02$ \citep{Federrath+10}. Notice that here we have converted the rms value of the skewness from 81 snapshots given in Table 1 of \citet{Federrath+10} 
to the uncertainty in the measured mean skewness  by dividing their reported rms value by $\sqrt{81}$.
This nearly lognormal shape of our composite PDF is consistent with the earlier observation that the effective driving of MC turbulence is 
more solenoidal than compressive.
 
The thin solid-line plot in Figure \ref{pdf} shows the average of the normalized PDFs for the clouds selected 
after gravity is included. Gravity gives rise to a power-law tail due to the formation of dense cores, 
as found in previous numerical studies \citep[e.g.][]{Scalo+98,Slyz+05pdf,Vazquez-Semadeni+08pdf,Kritsuk+11pdf,Collins+11,Collins+12,Federrath+Klessen13}. 
\citet{Kritsuk+11pdf} showed that, if a collapsing core has a density profile, $r^{-\alpha}$, it would contribute a power-law tail, $\rho^{-(1+3/\alpha)}$, 
to the distribution, $p_\rho$, of the density, $\rho$. In the inset of Figure \ref{pdf}, we show the overall probability 
distribution, $p_\rho$, of the density, $\rho$, for all the gas in the selected clouds in the presence of gravity. 
The right tail exhibits a $\rho^{-2.64}$ power law, corresponding to a $\rho (r) \propto r^{-1.83}$ density profile 
for collapsing cores.

\section{Discussion}

We have shown that the compressive ratio of SN-driven turbulence within MCs follows a broad lognormal distribution, with an average value 
$\langle \chi\rangle \approx \langle \chi_{\rm t}\rangle \approx 0.3$. Here we provide tentative arguments to explain why the mean value is lower than 
the equipartition one and why $ \chi$ exhibits a broad distribution.  

The first baroclinic effect discussed in section 3 is concerned with the effective SN driving through a pressure source term, and 
we have argued that the compressive ratio of the effective driving acceleration is likely close to the equipartition value of 0.5. 
The argument is supported by the fact that the relative directions of the pressure and density gradients in our simulated flow 
is roughly random. In Paper I, we computed  the compressive and solenoidal power spectra of the whole computational volume, 
$E_{\rm c}(k)$ and $E_{\rm s}(k)$. Inspection of Figure 8 in Paper I shows that $E_{\rm c}(k)/E_{\rm s}(k)\approx 0.5$ at 
the effective energy-injection scale, $L_{\rm in} \approx 70$ pc, consistent with the compressive ratio of the effective driving 
acceleration being close to 0.5. This shows that SN driving is not purely compressive, and that the average value of $\chi$ in 
MCs is expected to be no higher than 0.5.   

Furthermore, the compressive ratio computed from the solenoidal and compressive spectra drops rapidly with increasing 
wavenumber, $k$, or decreasing length scales (see Paper I). We argued that this rapid drop in $\chi$ (or steep decrease 
of the compressive spectrum with increasing $k$) is due to a second baroclinic effect, which preferentially converts compressive 
motions into solenoidal ones. This is the general baroclinic effect (more general than the first one from the SN pressure source) 
arising when the pressure and density gradients are misaligned (e.g., when a SN shock sweeps over a dense cloud).  As discussed 
in Paper I, due to its dependence on pressure and density gradients, the baroclinic effect is more efficient at small scales, meaning 
that it ``extracts" energy from compressive modes and converts it to solenoidal modes faster at smaller scales, causing the rapid drop
of the compressive spectrum toward larger $k$. This efficient conversion of compressive modes into solenoidal ones at small scales is 
supported by the finding that the baroclinic effect contributes to the  production of vorticity at a similar rate as vortex stretching (Kritsuk, 
private communication). Due to this baroclinic effect, $\chi$ is typically smaller than the equipartition value of 0.5 at inertial-range scales. 
The average value of the effective size of our cloud bounding boxes, $D_{\rm c}= 2\,R_{\rm c}$, where $R_{\rm c}$ is the effective radius 
defined in equation (\ref{eq_rc}), is 48.2 pc, just below the energy-injection scale of our simulated flow. Figure 8 of Paper I shows that a 
value of approximately 0.3 is consistent with the time-averaged value of $E_{\rm c}(k)/E_{\rm s}(k)$ at a scale of approximately 48 pc. 

The broad distribution of $\chi$ can be understood by considering the amplitudes, $\Delta v_{\rm s}(D_c)$ and $\Delta v_{\rm c}(D_c)$, 
of the solenoidal and compressive velocities at the cloud size. If $D_c$ is close to the energy injection scale, $L_{\rm in}$, the distributions 
of $\Delta v_{\rm s}(D_c)$ and $\Delta v_{\rm c}(D_c)$ are roughly Gaussian. Since the compressive power ratio is related to the ratio 
of $\Delta v_{\rm c}(D_c) /\Delta v_{\rm s}(D_c)$, and the ratio of two Gaussian variables has a distribution much broader than 
Gaussian\footnote{The distribution of the ratio of two Gaussian variables is a Lorentz distribution, which is indeed very broad.}, 
one expects $\chi$ to show a broad, non-Gaussian distribution, as found in section 5.   
Furthermore, at scales $D_c$ below $L_{\rm in}$, the solenoidal and compressive velocity amplitudes, $\Delta v_{\rm s}(D_{c})$ 
and $\Delta v_{\rm c}(D_{c})$, would become non-Gaussian due to turbulent intermittency. This tends to make the distribution of 
the ratio $\Delta v_{\rm c}(D_{c})/\Delta v_{\rm s}(D_{c})$ at small scales (and hence the distribution of $\chi$ for smaller clouds) 
even more non-Gaussian with fatter tails. \\

\section{Conclusions}

In this work, we have analyzed the SN-driven simulation of interstellar turbulence presented in Paper I, focusing on the statistics
of the compressive ratio of the turbulence within MCs. Our main results are as follows:

\begin{enumerate}

\item The estimated compressive ratio of SN-driven turbulence within MCs follows a broad lognormal distribution, 
with the average $\langle \chi_{\rm t}\rangle=0.28\pm 0.17$ in the case with gravity, and $\langle \chi_{\rm t}\rangle=0.30\pm0.18$ 
without gravity, comparable to the value found in isothermal simulations with random solenoidal driving and sonic or Alfv\'{e}nic Mach 
numbers $\approx 3$, and significantly lower than in isothermal simulations with purely compressive driving.
Self-gravity does not affect the compressive ratio significantly. 
 
\item The mean expansion or contraction velocity, $V_{\rm e}$,  in the clouds is, on average, only a small fraction of the total rms velocity,
with $\langle |V_{\rm e}|/\sigma_{\rm v}\rangle = 0.15$ for the contraction velocity of both clouds with and without 
self-gravity. Thus, even in the presence of self-gravity, MCs do not collapse as a whole. However, 
the fraction of clouds that are contracting grows from 45\% to 62\% after gravity is included.

\item The cloud solid-body rotation velocity is larger than the mean radial velocity, but still a small fraction of the total rms velocity on average, 
with $\langle |V_{\rm r}|/\sigma_{\rm v}\rangle = 0.28$ and 0.27 for clouds before and after the inclusion of self-gravity. Thus, 
cloud rotation is not significantly affected by gravity either.

\item The amplitude of the density fluctuations in MCs follows approximately a similar $\sigma_{\rm s}$-- ${\cal M}$ relation as in idealized simulations 
with random driving, but with a very large scatter around the mean $\sigma_{\rm s}$-- ${\cal M}$ curve. 

\item Although significant deviations may exist for individual MCs, the composite gas density PDF obtained by 
the combination of the normalized PDFs of all clouds, in the absence of gravity, is very well described by a lognormal distribution, 
over nearly seven orders of magnitude for the probability at the high-density tail. 
Once gravity is included in the simulation, the PDF develops a power-law high-density tail, 
$\sim \rho^{-2.6}$, due to the presence of collapsing cores. 

\end{enumerate} 

Our results demonstrate that numerical studies of MC turbulence on small to intermediate scales, which attempt to imitate the effect of the ISM turbulent 
cascade by using a large-scale random acceleration, should adopt a driving scheme resulting in values of $\chi_{\rm t}$ consistent with its 
distribution derived here, assuming SN explosions are the main driving mechanism of the ISM turbulence. 
This distribution should also be accounted for when modeling star formation with statistics of supersonic turbulence.

\acknowledgements

We acknowledge useful comments by the anonymous referee and discussions with Alexei Kritsuk that helped us improve the 
manuscript. Computing resources for this work were provided by the NASA High-End Computing (HEC) Program through 
the NASA Advanced Supercomputing (NAS) Division at Ames Research Center, by PRACE through a Tier-0 award providing 
us access to the computing resource SuperMUC based in Germany at the Leibniz Supercomputing Center, and by the Port 
d'Informaci\'{o} Cient\'{i}fica (PIC), Spain, maintained by a collaboration of the Institut de F\'{i}sica d'Altes Energies (IFAE) 
and the Centro de Investigaciones Energ\'{e}ticas, Medioambientales y Tecnol\'{o}gicas (CIEMAT). PP acknowledges support by 
the Spanish MINECO under project AYA2014-57134-P. TH is supported by a Sapere Aude Starting Grant from The Danish 
Council for Independent Research. Research at Centre for Star and Planet Formation was funded by the Danish National 
Research Foundation and the University of Copenhagen's programme of excellence.

%\bibliographystyle{apj}
%\bibliography{apj-jour.bib,MC,padoan,nsf08,paper_SN}

\end{document}